\documentclass{jalc}

\usepackage{graphicx}
\usepackage{enumerate}
\usepackage{amssymb}
\usepackage{color}

\begin{document}%
     %
\title{From Helmut J\"urgensen's Former Students: The Game of Informatics Research}
     %

\runningtitle{From Helmut J\"urgensen's Former Students: The Game of Informatics Research}
     
\runningauthors{\textsc{M.~Daley}, \textsc{M.~Eramian}, \textsc{I.~McQuillan}, \textsc{C.~Power}}
     

\author[UWO]{Mark Daley}
\address[UWO]{Department of Computer Science,
  University of Western Ontario\\
  Middlesex College,
  London, ON,
  Canada\\
  \email{mark.daley@uwo.ca}
}

\author[UofS]{Mark Eramian}
\address[UofS]{Department of Computer Science,
  University of Saskatchewan,\\
  110 Science Place,
  Saskatoon, SK,
  Canada\\
  \email{eramian@cs.usask.ca}
  \email{mcquillan@cs.usask.ca}
}

\author[YORK]{Christopher Power}
\address[YORK]{Department of Computer Science,
  University of York,\\
  Heslington, York, UK\\
  \email{christopher.power@york.ac.uk}
}

\author[UofS,IAN]{Ian McQuillan}
\thanks[IAN]{Supported in part by a grant from NSERC.}
\thanks{Published at Journal of Languages, Automata and Combinatorics, 23, 2018, 127--141 DOI: 10.25596/jalc-2018-127}

\maketitle

%
\begin{abstract}
Personal reflections are given on being students of Helmut 
J\"urgensen. Then, we attempt to address his hypothesis that
informatics follows trend-like behaviours through the use
of a content analysis of university job advertisements, and then
\textit{via} simulation techniques from the area of quantitative economics.
\end{abstract}

%
%
%
\section{Personal Reflections}\label{s:in}

The four authors of this paper were students of Helmut J\"urgensen, with the latter three supervised by Helmut for all of our graduate studies.
We start with personal reflections on this experience.

\subsection{A Note from Mark Daley}

I recall sitting at a desk in the rotunda outside Helmut's office reading Saunders Mac Lane's \textit{Categories for the Working Mathematician}. Helmut stepped out of his office, regarded me for a moment, and then said: ``It is certainly beautiful, but one must be careful not to lose one's entire life to it.''  While I was not officially a student of Helmut's, I consider him my adoptive second Doktorvater nonetheless, and this interaction is illustrative of the thoughtful mentoring I received from him. 

Helmut and I share many interests: computability, information, art, music, language, semigroups, typesetting, history, computing technology, and more; I have fond memories of long conversations on these topics with Helmut, and his students, from my time as a graduate student. Amongst all that I learned from Helmut, there are two lessons that I adopted as guiding personal philosophies and to which I believe I owe much of the good fortune I have had in my own career. The first: \textit{Formalize everything.} When dealing with formal systems, it is too easy to fool oneself to incorrect conclusions with ad hoc informal reasoning. When I collaborated with him, Helmut always insisted on complete rigour in everything we did, and the end product was always stronger for it. The second, and most important: \textit{Be fearlessly interdisciplinary}. The fluidity with which Helmut moved between research questions and disciplines solidified in my mind an exemplar of the type of scholar I aspired --- and still do aspire --- to be.

\subsection{A Note from Mark Eramian}

When I think of my time as Helmut's student I remember the deep discussions, the unwavering confidence he had in me even when I didn't feel confident in myself, the summer backyard BBQ's Helmut held for his students at his house annually, and, of course, the stories.  I had the fortune of knowing Helmut before I even enrolled in university.  After my first undergraduate year, I approached him to ask if he knew of any professors who were seeking summer research assistants.  He asked around and, finding none, offered to hire me himself.  That summer project was my first foray into research, and probably why I stuck with it.  I recall spending that summer reading Helmut's Pascal code (ported from Algol 60 of course) with a German-English dictionary at my side to decipher the variable names.  Helmut kept me engaged in research throughout my undergraduate career with many interesting projects and in graduate school allowed me to combine my interests in digital image processing and theoretical computer science in a way that probably not many could.  He taught me the thrill of success and that failure is normal.   He cultivated in me a profound love of \LaTeX, a pedantic insistence for typesetting perfection, and a crippling sensitivity to poor typographic kerning.  

In supervising my own students I look to Helmut's example of patience and supportiveness.  Working with Helmut was one of the most inspirational and fun times of my life, and I am thankful for the experience.  I am thrilled and proud to collaborate with my contemporaries on this paper in Helmut's honour.

\subsection{A Note from Ian McQuillan}
I have very fond memories of my time as a graduate student working under the supervision of Helmut J\"urgensen. I first met Helmut while taking an advanced course on automata and formal language theory as an undergraduate student. I was excited by the foundational aspect of his emphasis; we learned about G\"odel's incompleteness and completeness theorems, and also primitive recursive functions. Despite being a ``less than stellar'' undergraduate student, Helmut recognized my enthusiasm and potential, and agreed to supervise me for graduate school. Upon starting, he suggested that I work on the topic of synchronized grammars, where there was quite a bit of ``low hanging fruit'' for a young graduate student to make quick progress and build up their confidence, an approach I now try to take with my own graduate students. I am almost certain that I would not have ended up working in academia had it not been for Helmut's supervision. 

He created an excellent environment for graduate students. He would frequently enter the student offices telling stories about leaders in the field, computer science through different eras, and interesting research problems. The Department of Computer Science at the University of Western Ontario excelled in theoretical computer science and natural computing, which would become my main research interests. There were lots of opportunities for discussion and collaboration, and Helmut left research incredibly open-ended, where I could pursue any topic that struck my interests. I also try to duplicate this environment and approach with my students, wherever possible. Many of Helmut's students ended up being lifelong professors, in a diverse set of areas. Even the authors of this article who were graduate students at roughly the same time work are now professors in many areas, such as image processing and medical image analysis, data analytics and information visualization, accessibility for people with disabilities, computational neuroscience, theoretical computer science, and natural computing. This diversity speaks to his approach.

\subsection{A Note from Christopher Power}

My warmest memories of working with Helmut are of him encouraging me to pursue a research career at the end of my first year of my doctoral studies, when I was lost in a sea of proofs and theories, uninspired, and considering leaving.  He kindly took me aside, and said ``There are some computer scientists who are inspired by the infinite beauty of mathematics.  There are others who are fascinated by creating new and wondrous things.  You are that second kind, embrace it, and be happy.''  Over a decade later, I still start my programming lectures with that story.  Helmut encouraged me to work with people with disabilities, not only because it is a noble thing to do, but also because of the challenge of thinking outside the traditional boundaries of software design.  He connected me with people such as the authors of this paper, who have been lifelong friends and colleagues.  He introduced me to senior researchers in Canada and Europe and, as I found out years later, cold-called my first boss to tell her the potential I had, and what an asset I would be to her team.  I can safely say, my life and my career would have been far less interesting and meaningful if not for his friendship and guidance.

In meetings, after sharing stories of new discoveries, bits of gossip from other groups around the world, and lamenting languages he had forgotten over the years, Helmut would always say to me ``Now go away, and come back and tell me something interesting.''  With my own students, I tell them the same thing, for that is possibly the best encouragement a student can have: knowing that they have something important to say and that someone will listen.  Hopefully, my friends and I have honoured his sentiment in this paper and have found something he will think is interesting.

\section{The Game of Informatics Research}

Helmut would frequently enter the student lab where we worked and tell stories about researchers and academia in general. When we were reminiscing about this, we all recalled one particular analogy that he made relating computer science and informatics research to the board game Clue.  Clue (Parker Borthers, 1949) is a 3-6 player game in which players control characters which they move from room to room in a house attempting to solve a murder by guessing where, by whom and with what weapon the murder was committed, eventually coming to the solution by a process of elimination.

\begin{quote}
{\it ``Informatics in those growing-up years was a field of fast-moving focus. Informatics then and even today reminds me of the game of Clue --- or the corresponding movie ---, in which a Mr.\ Body has been murdered and the remaining group of people investigates by running around in a panic from one room to another and back again --- occasionally leaving someone behind. Once the easy problems of a field have been solved, most people move on to the next fashionable field, possibly even proclaiming the old field dead.''}

\noindent
--- H. J\"urgensen, ``People and Ideas in Theoretical Computer Science'' \cite{people99}.
\end{quote}

In relating this to us in person, Helmut would also add additional flavour to the analogy. He would say that several very clever people would enter a room, say the conservatory, and start asking some clever questions. Then, upon everyone else in the game hearing those questions, they would all run into the conservatory and start asking very similar questions, with small variations.  The rope is replaced with the wrench.  Miss Scarlet is replaced with Colonel Mustard.  Eventually, you find that everyone is asking the same questions.
Then, the clever people who first were in the room, move onto the dining room and begin asking new questions.  The crowd then races behind them into the dining room and begins asking the same questions they asked in the conservatory, because it is easy and comfortable.
However, there are some hard questions left in the conservatory, and some very clever people will stay behind and work to find and answer those questions.
Helmut would say: ``Your goal as a researcher is to be the first person in the room, or the last person in the room.''

We wanted to test the hypothesis that informatics research does indeed largely follow this trend-like behaviour. 

\section{\emph{Clue}-like Behaviour in Hiring}

When considering the game of \emph{Clue} in informatics research, it is first worth considering whether this type of behaviour is actually prevalent in the research community.  Certainly, there are trends that happen in funding regimes world-wide.  It has been the experience of the authors that funding agencies such as NSERC in Canada, NSF in the USA, RCUK in the United Kingdom, or framework calls from the European Commission drive into different areas over time. We proceed on this assumption for which there is recent evidence.  Indeed, large commitments have been made over the last decade sequentially to bioinformatics, high-performance computing, and games.  As we move deeper into the 21\textsuperscript{st} century, and governments become more driven by a need to meet national industrial objectives, it is likely that these surges of funding to specific areas will continue or become more pronounced.    However, do the institutions respond in kind?  Do the hiring practices focus on broader areas, or do they specialize in ways that are comparable to the way funding is targeted at particular subareas?  Anecdotally, anyone who follows job advertisements for a period of more than a couple of years will think there is a similar trend; however, to our knowledge, this question has never been empirically investigated.  In this section, we undertake a qualitative study of job advertisements to answer the question as to whether trends in hiring seem to support the notion of a \emph{Clue}-like game taking place in the Computer Science academy.

In order to answer this question, we undertook a content analysis of a set of job advertisements from Canada and the United States.  Content analysis is a means of making inferences about text that are both reliable and replicable by researchers~\cite{Krippendorff:2012}.  This is usually done by taking a collection of written source texts and producing markup codes for particular types of items contained within.  In this case, we are interested in identifying, coding, and making inferences about the types of researchers that institutions hire over time.  We are interested in identifying whether there is a prevalence of institutions moving rapidly into new ``rooms'' to chase research money and/or find trendy questions to ask, only to move on again soon thereafter.  

\subsection{Source Selection}
We identified the Computer Research News (CRN) archive produced by the Computer Research Association (CRA) as an ideal set of documents for the purpose of answering these questions.  The CRN is a broad-reaching publication, with a wide readership by North American academics and beyond, that has been consistently published over the last 25 years, originally 5 times a year and later moving to 10 times per year in 2012.  

\begin{figure}
\includegraphics[width=\textwidth]{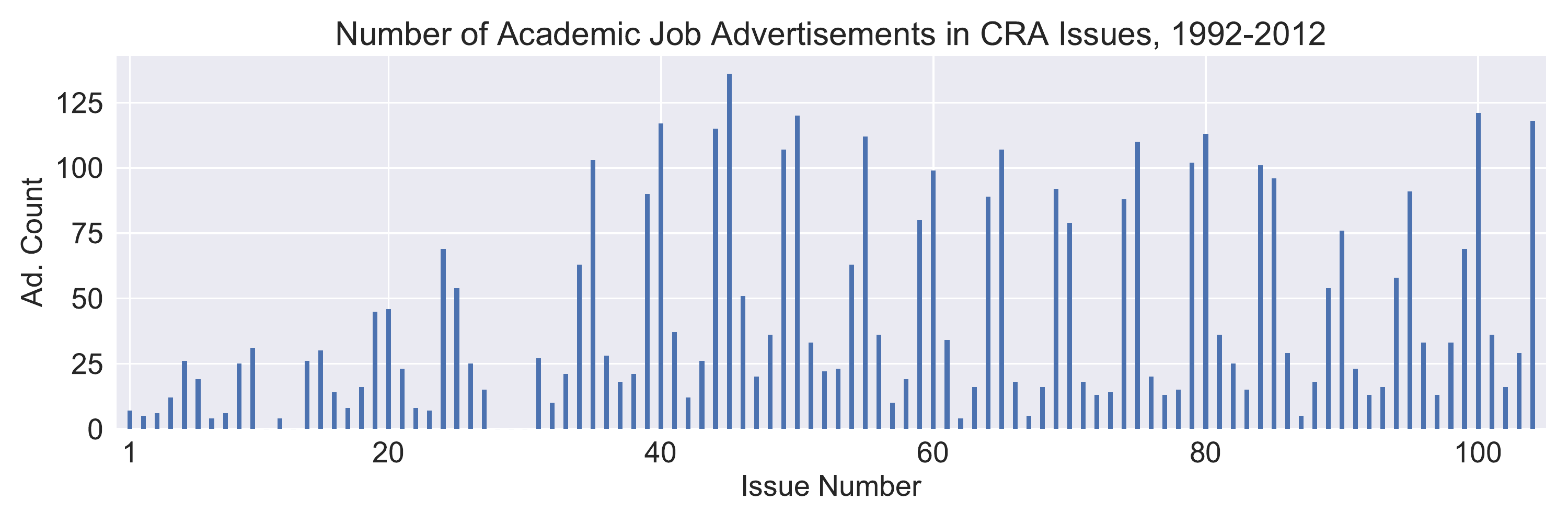}
\caption{Number of academic job advertisements per CRA issue over time between 1992 and 2012.  The hiring cycle is evident from the peaks of advertisements in the November and January issues.  Issues with a count of zero are missing from the archive or were corrupted and unreadable.}\label{fig:adcounts}
\end{figure}

We selected a 20 year period from 1992 through to 2012 which captures a major period of growth and diversity for the informatics discipline.  Each publication has a set of job opportunities advertised as part of its main content, with the number of advertisements per issue ranging from 5 to 135.  In terms of distribution, in general the first publication of each year (January) and the fifth publication (November) contained the largest proportions of job opportunities, corresponding with typical hiring start dates of July or September at most institutions.   Figure \ref{fig:adcounts} shows the number of advertisements per issue over time.  The number of ads increased a lot over the first 34 issues in the dataset, corresponding to the period from 1992 to 1998, then became fairly steady.  The aforementioned hiring cycle is also evident from the peaks and valleys.

\subsection{Units of Analysis}
We chose to examine each individual advertisement and code the topics of research candidates should pursue in order to be desirable or preferable for hiring.  We included in our data collection all advertisements for academic research positions  and industrial research positions, including visiting professor positions if they required a research component.  Advertisements for postdoctoral fellows, department heads, administrative/director positions, teaching-only positions, and clearly non-research industrial positions were excluded.  In all, 3248 of 4276 ads met our inclusion criteria.

In place of having a completely open and emergent classification scheme, we examined the first issue from each year over the 20 year period, and captured the most common areas that were advertised for.  This produced an initial list of 27 areas.  It was noted during this process that there was a wide variety of different terms used by the community for describing identical areas of study.  We also included a category \emph{Other} to capture those items that did not seem to fit one of the original categories.  When using this code, we recorded the exact words from the advertisement for purposes of later review.  After coding was complete, this \emph{Other} category and its collected terms were examined for common terms.  Some terms were fit into existing categories (e.g. multicore was included in \emph{parallel and distributed computing}).  For those areas remaining in the \emph{Other}, category, they were merged where there were recording differences (e.g. \emph{e commerce} vs. \emph{e-commerce}) or where they described the same area in different ways (e.g. \emph{multi-agent systems} and \emph{agent-based systems}).   Coding to determine the list of categories as well as subsequent coding of the remainder of the dataset was performed manually by three of the authors (M.E., I.M., and C.P.).


The complete list of categories used in the coding scheme were: 
Open (no specific specialization was requested), 
Experimental Computer Science, 
Applied Computer Science, 
AI/Machine Learning (including related areas such as data mining), 
Big Data, 
Bioinformatics (including computational biology), 
Computer Architecture (including all hardware/computer engineering areas), 
Computer Graphics (including multimedia), 
Computer Vision (including image processing), 
Data Visualization, 
Databases, 
Games, 
High-Performance Computation, 
Human-Computer Interaction, 
Mobile Computing (including pervasive and ubiquitous computing), 
Modelling and Simulation (including operations research), 
Networks, 
Operating Systems, 
Parallel and Distributed Computing (including cloud computing), 
Programming Languages (including compilers), 
Scientific Computing (including computational science), 
Security (including cryptography, information assurance), 
Social Computing, 
Software Engineering, 
Theoretical Computer Science (including algorithms, data structures, language theory), 
Web Technologies, 
and the aforementioned Other category.

\subsection{Reliability}
We undertook an inter-coder reliability task to ensure that the coding of the data was done in a consistent, repeatable, and valid way.  Two researchers were given identical sets of 40 job advertisements selected from across the sample and across the codes.  Each researcher independently coded the sample using the coding scheme, and the sets were brought together for comparison.  To determine inter-coder agreement we used Cohen's kappa, defined as having $p_o$ as the relative observed agreement among coders, and $p_e$ is the probability of chance agreement by the coders.  

$$\kappa = \frac{p_o-p_e}{1-p_e}$$

With $\kappa=1$ representing perfect agreement, the coders of the job advertisements achieved $\kappa=0.83$.  When examining the results, many of the errors came from one individual coding two or more items within job advertisement separately, while the other merged these items into a single code.  When removing these duplication errors, the inter-coder agreement was $\kappa=0.96$, well above the threshold for having high confidence that the coding of the advertisements was done reliably.

\subsection{Results}

\begin{figure}[h]
\begin{center}
\includegraphics[width=\textwidth]{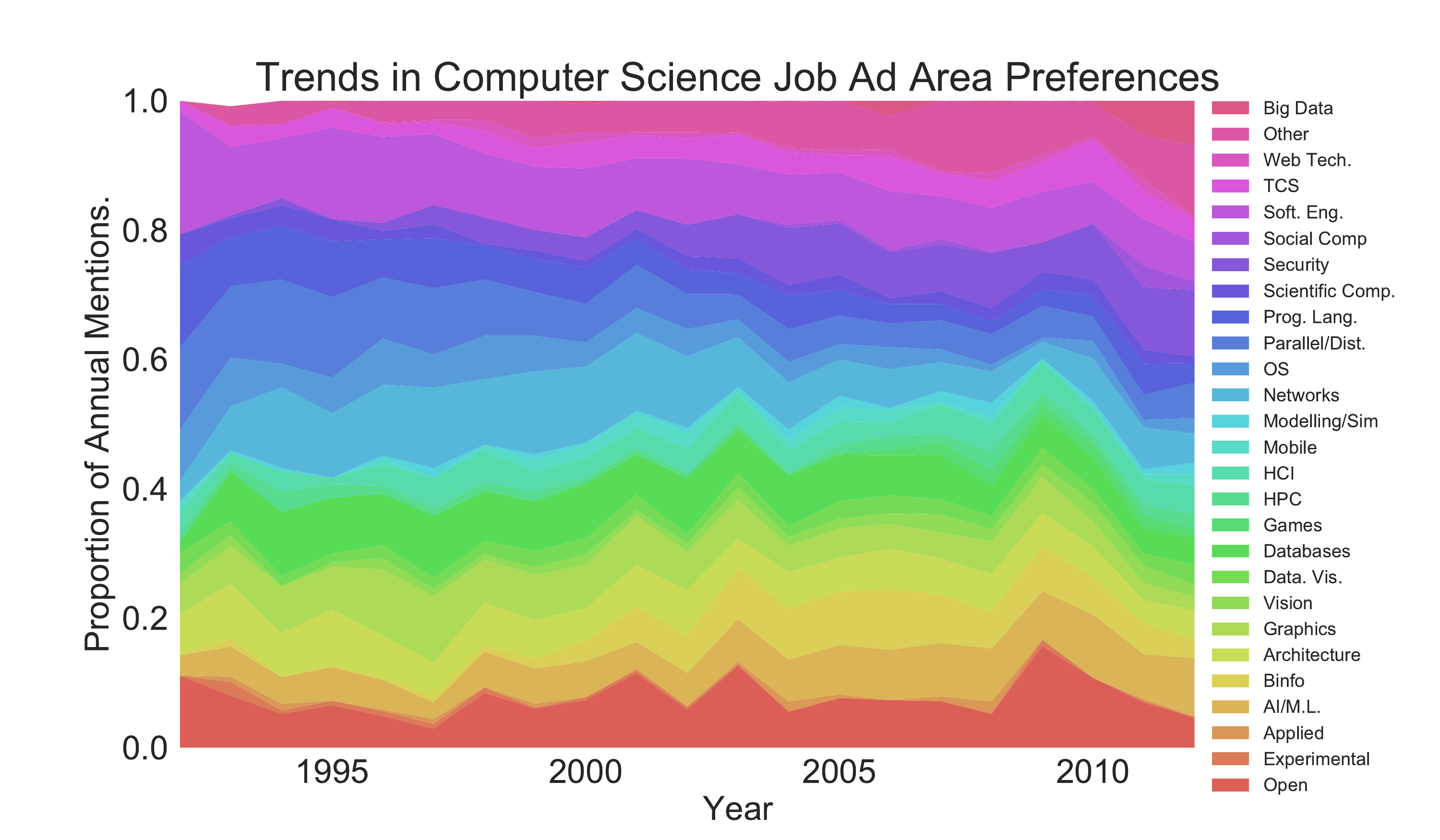}
\end{center}
\caption{Overall results showing academic ad preferred area trends.  The width of an area $i$'s band at year $j$ represents $p(i,j)$.}
\label{fig:overallresults}
\end{figure}

The recorded data were summarized in a matrix consisting of a row for each of the 27 topic areas, and a column for each year.  The entry in row $i$, column $j$ was the proportion of areas mentioned in year $j$ that were area $i$, denoted $p(i,j)$.  Formally, let $A_j(k)$ be the $k$-th ad in year $j$, $N(A_j(k))$ be the number of uniquely coded areas mentioned in an ad, and $N_j = \sum_k N(A_j(k))$ be the total number of topics mentioned in ads in year $j$.  Then
$$
p(i,j) = \frac{|\{ A_j(k) \mid \text{$A_j(k)$ mentions topic $i$}\}|}{N_j}.
$$
To get a visualization of the entire dataset, each row of the matrix $p$ was plotted in an area line graph, shown in Figure \ref{fig:overallresults}.  

\begin{figure}[t]
\begin{center}
\begin{tabular}{cc}
\includegraphics[width=.45\textwidth]{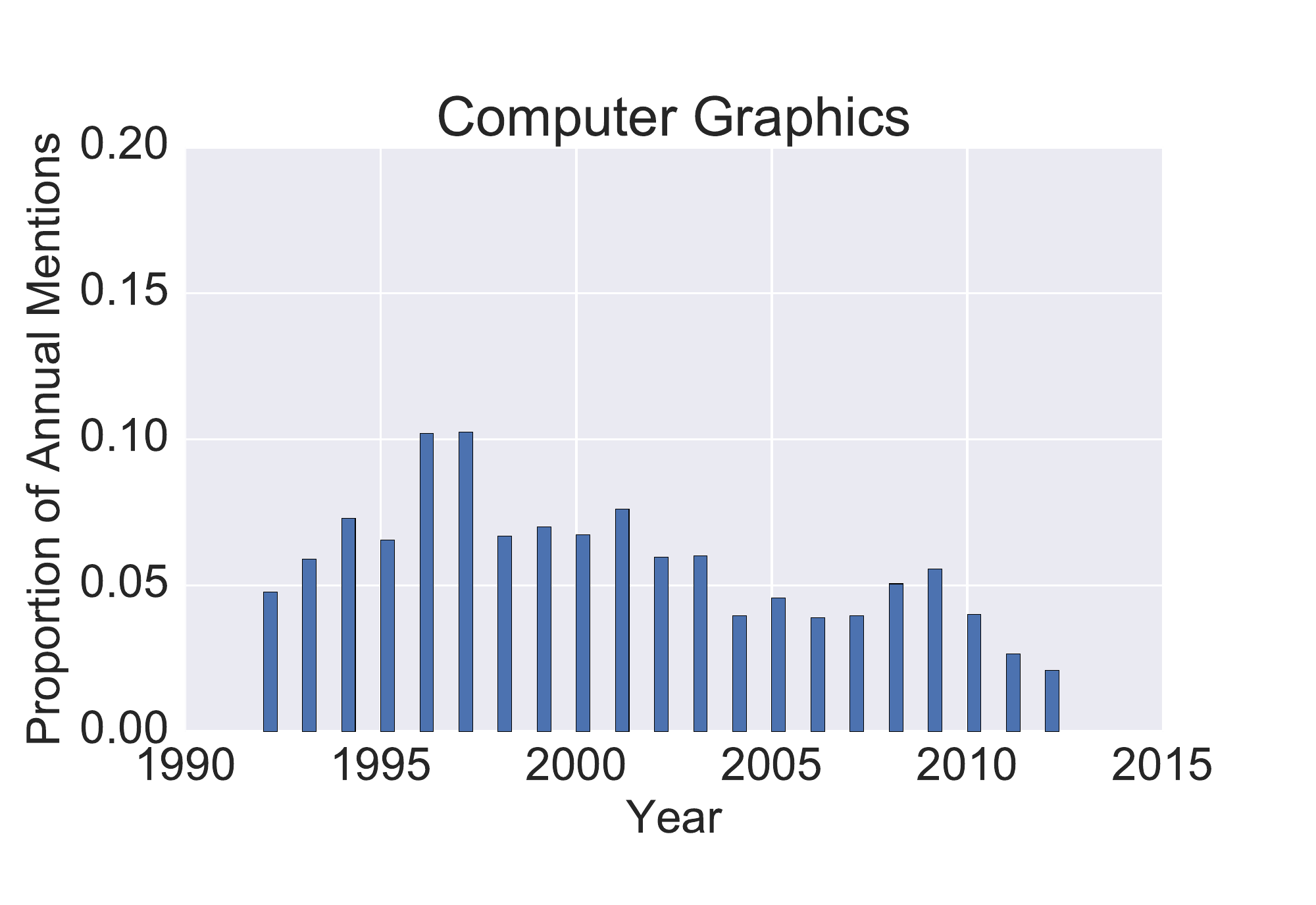}&
\includegraphics[width=.45\textwidth]{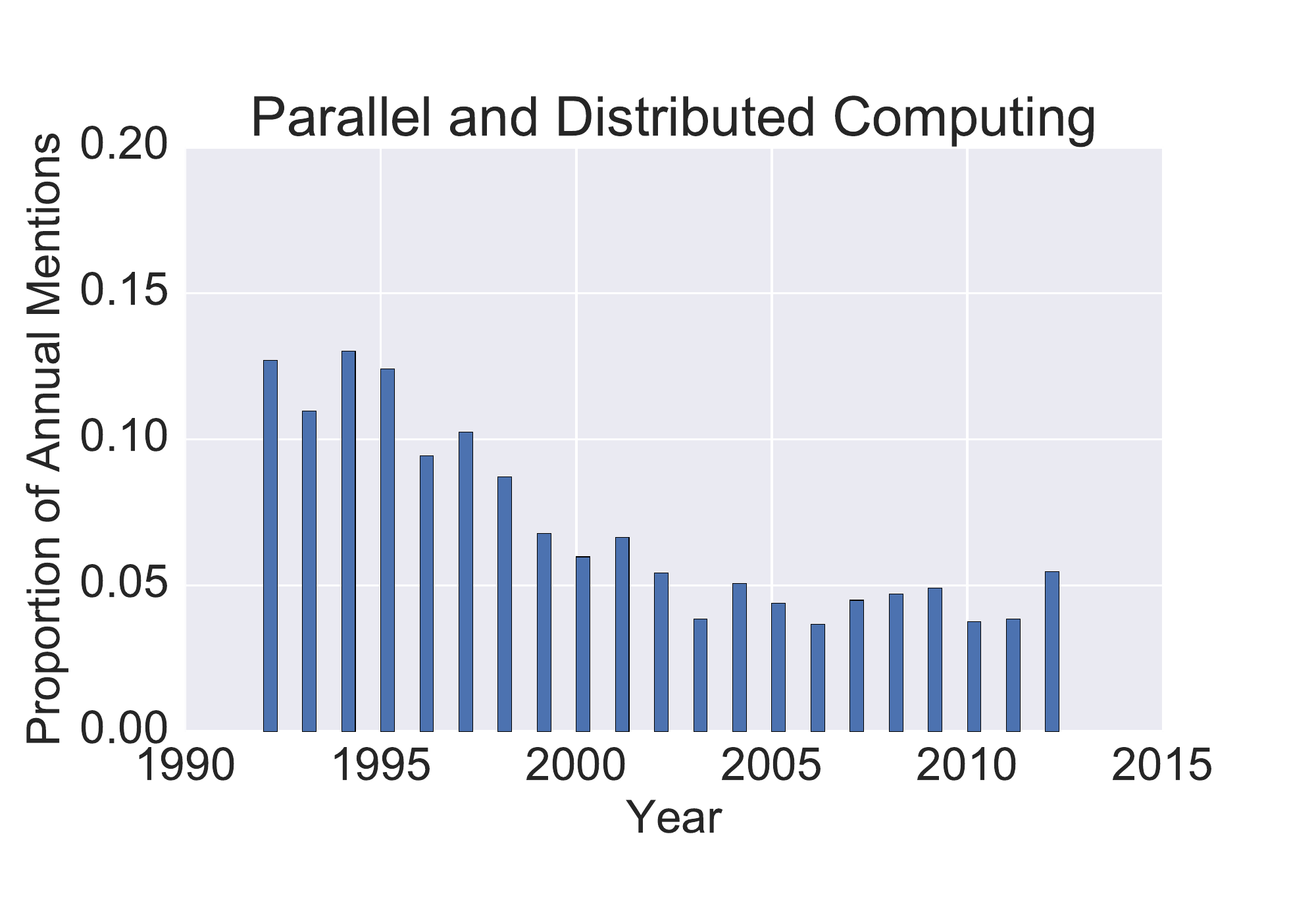}\\
\includegraphics[width=.45\textwidth]{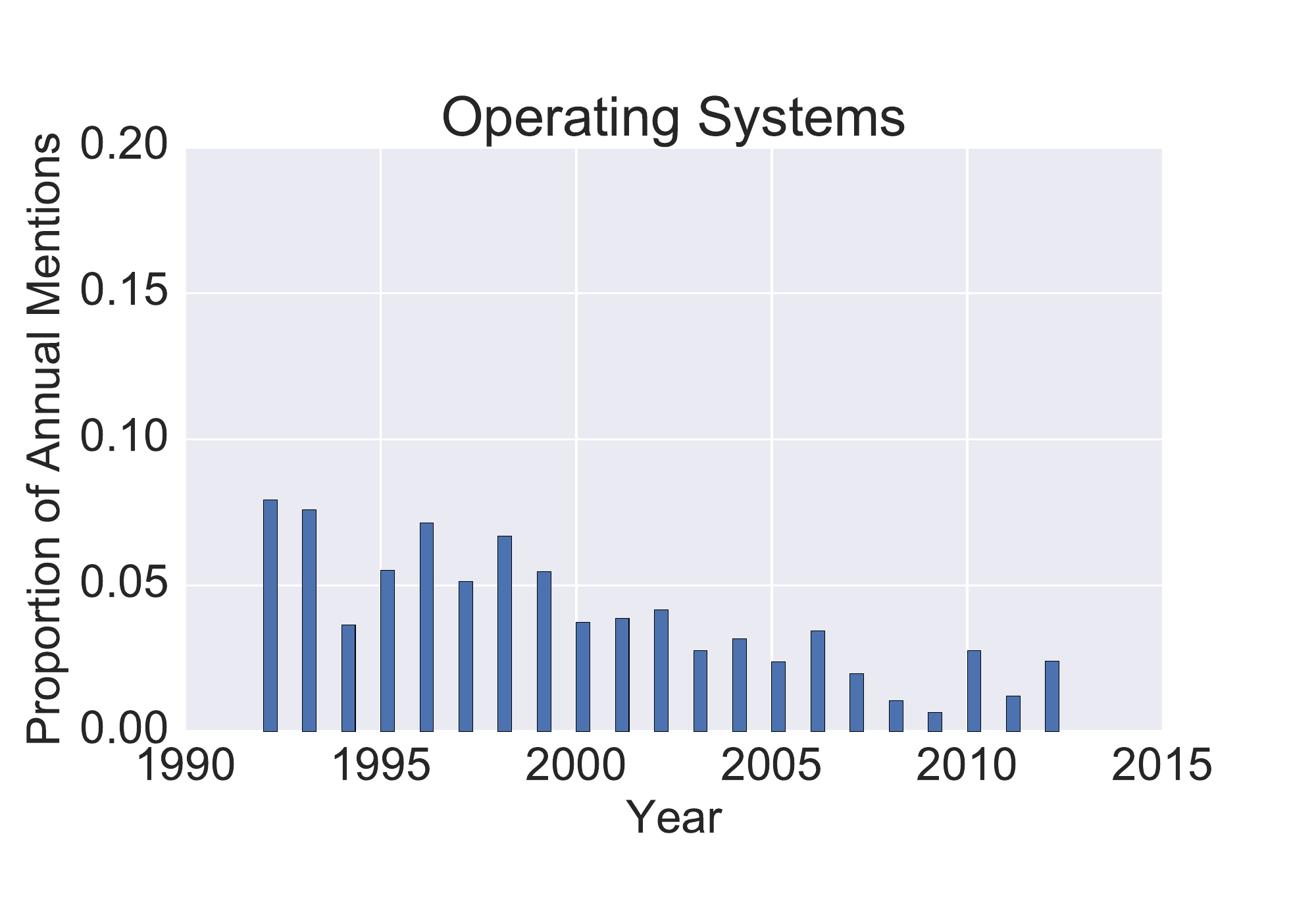}&
\includegraphics[width=.45\textwidth]{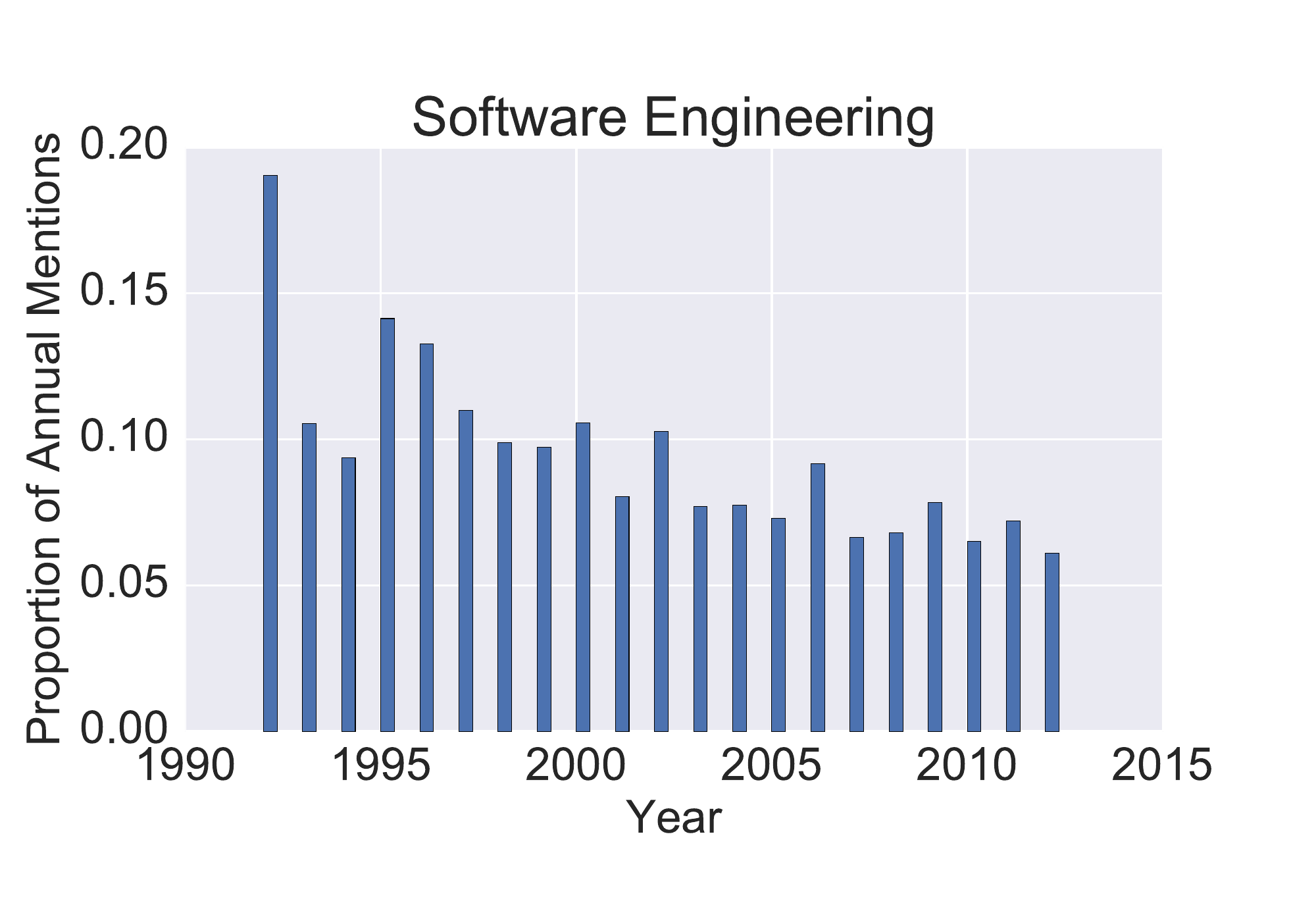}\\
\includegraphics[width=.45\textwidth]{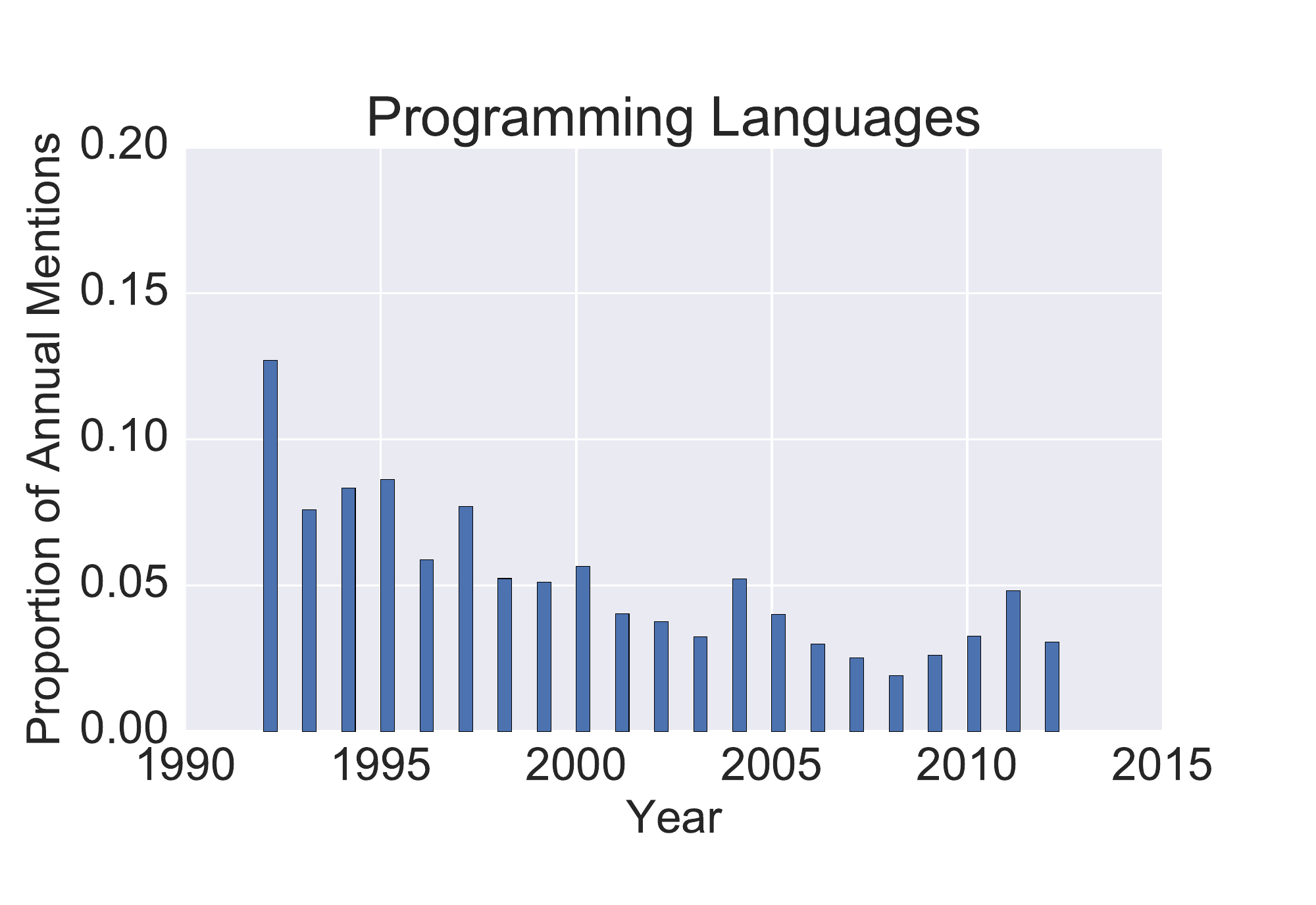}&
\includegraphics[width=.45\textwidth]{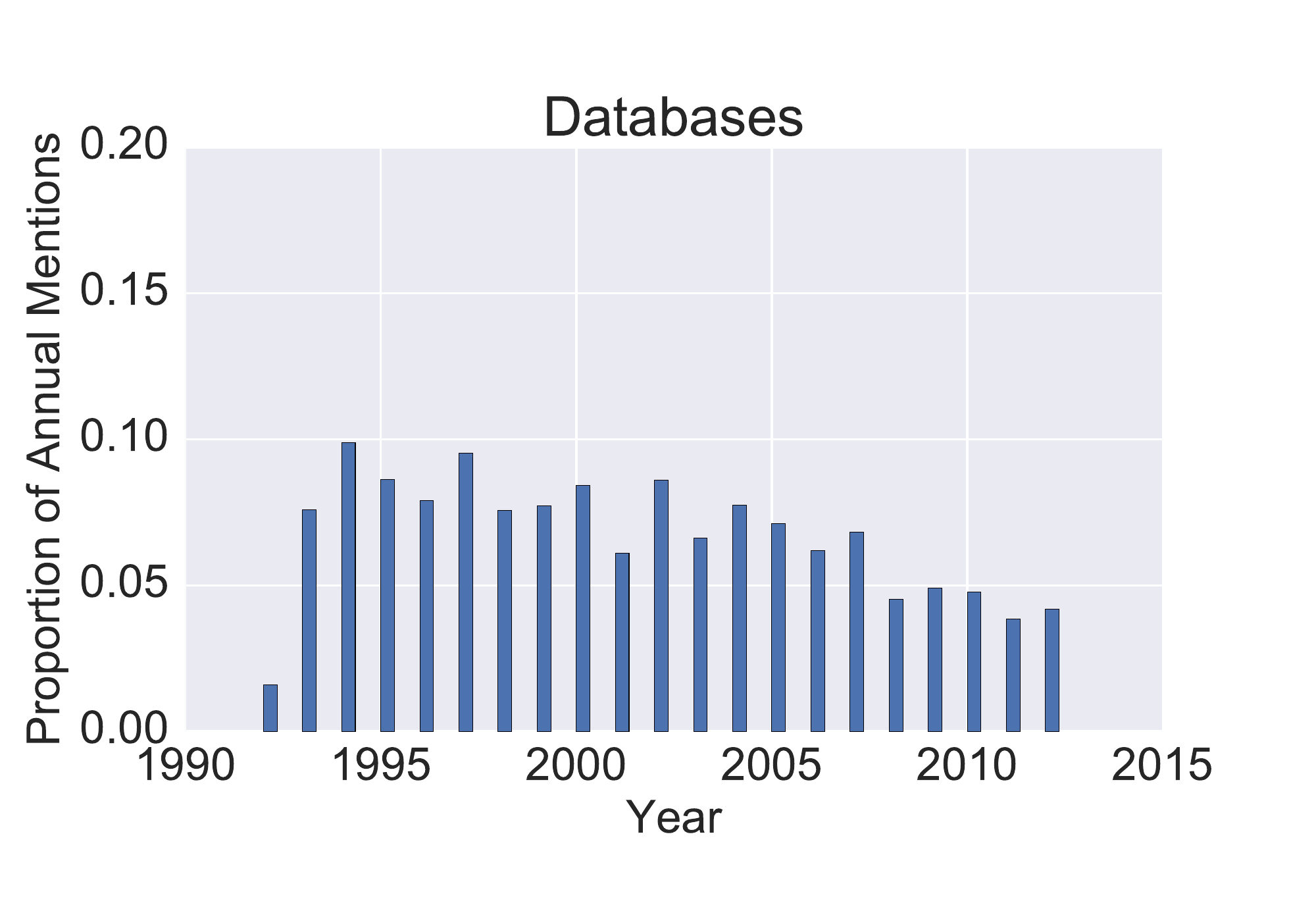}
\end{tabular}
\end{center}
\caption{Yearly proportion advertisements mentioning areas of Computer Science and Informatics that appear to have peaked around the beginning of our data set and are declining steadily.}\label{fig:declining}
\end{figure}

\begin{figure}[t]
\begin{center}
\begin{tabular}{cc}
\includegraphics[width=.45\textwidth]{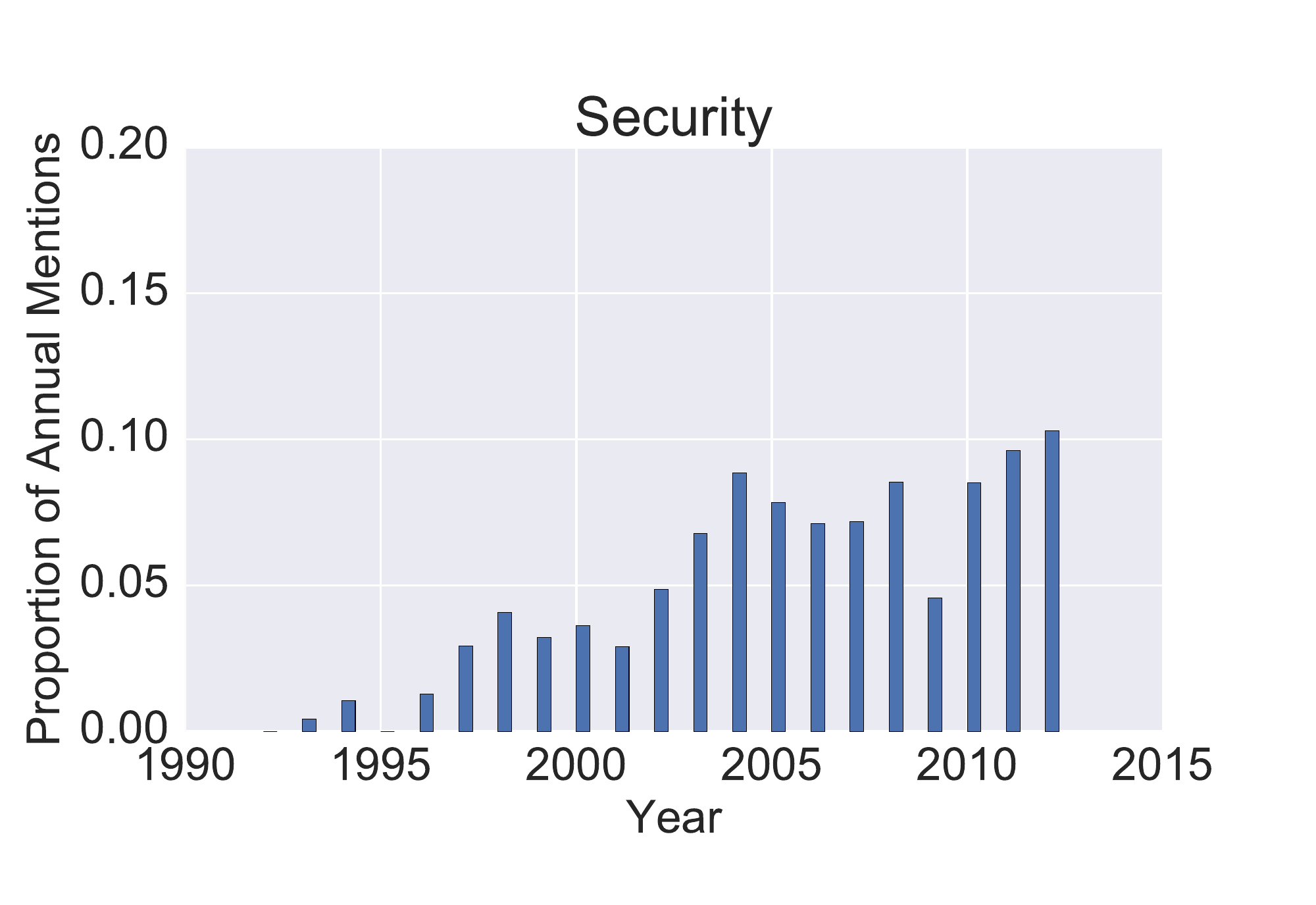}&
\includegraphics[width=.45\textwidth]{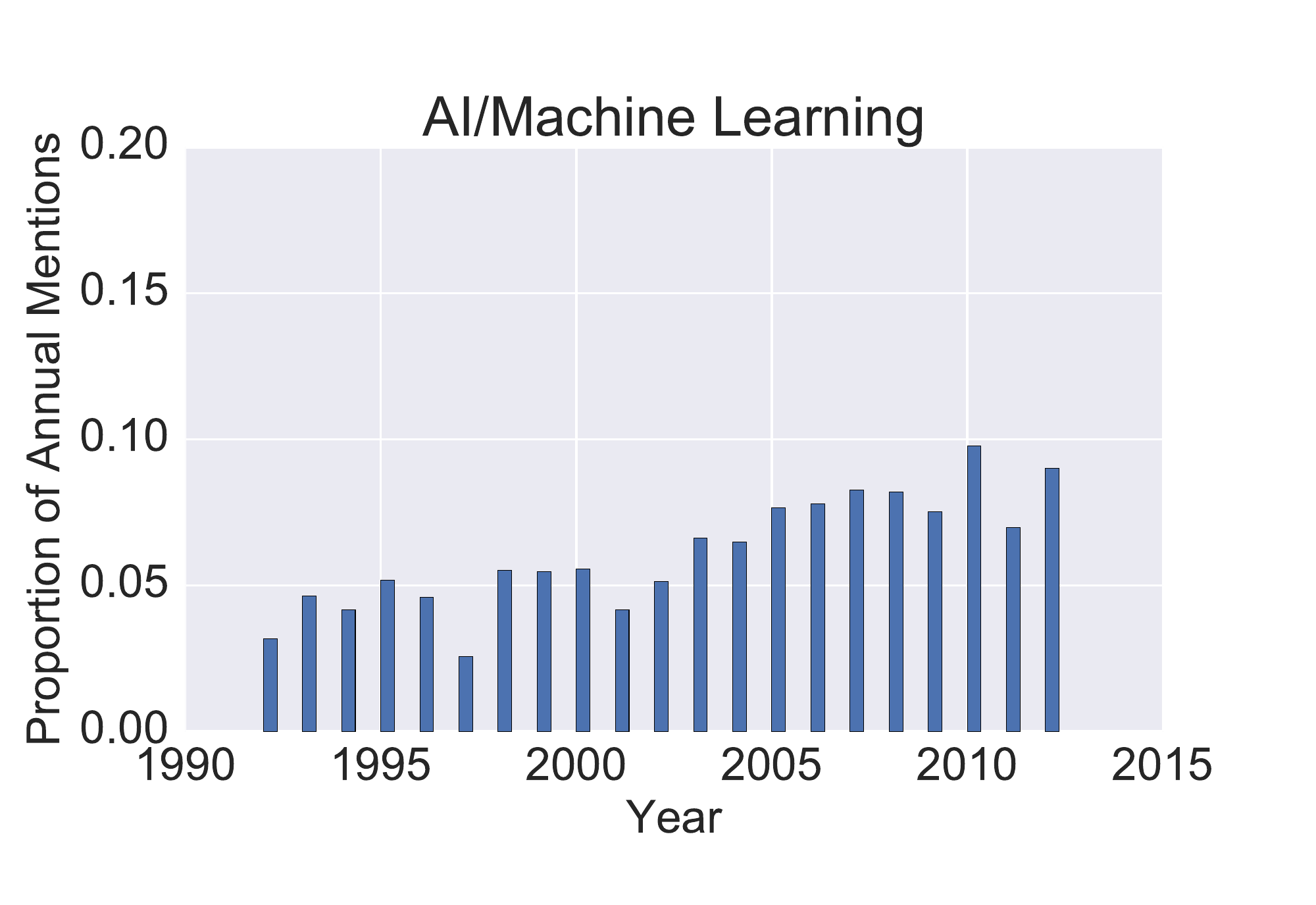}\\
\includegraphics[width=.45\textwidth]{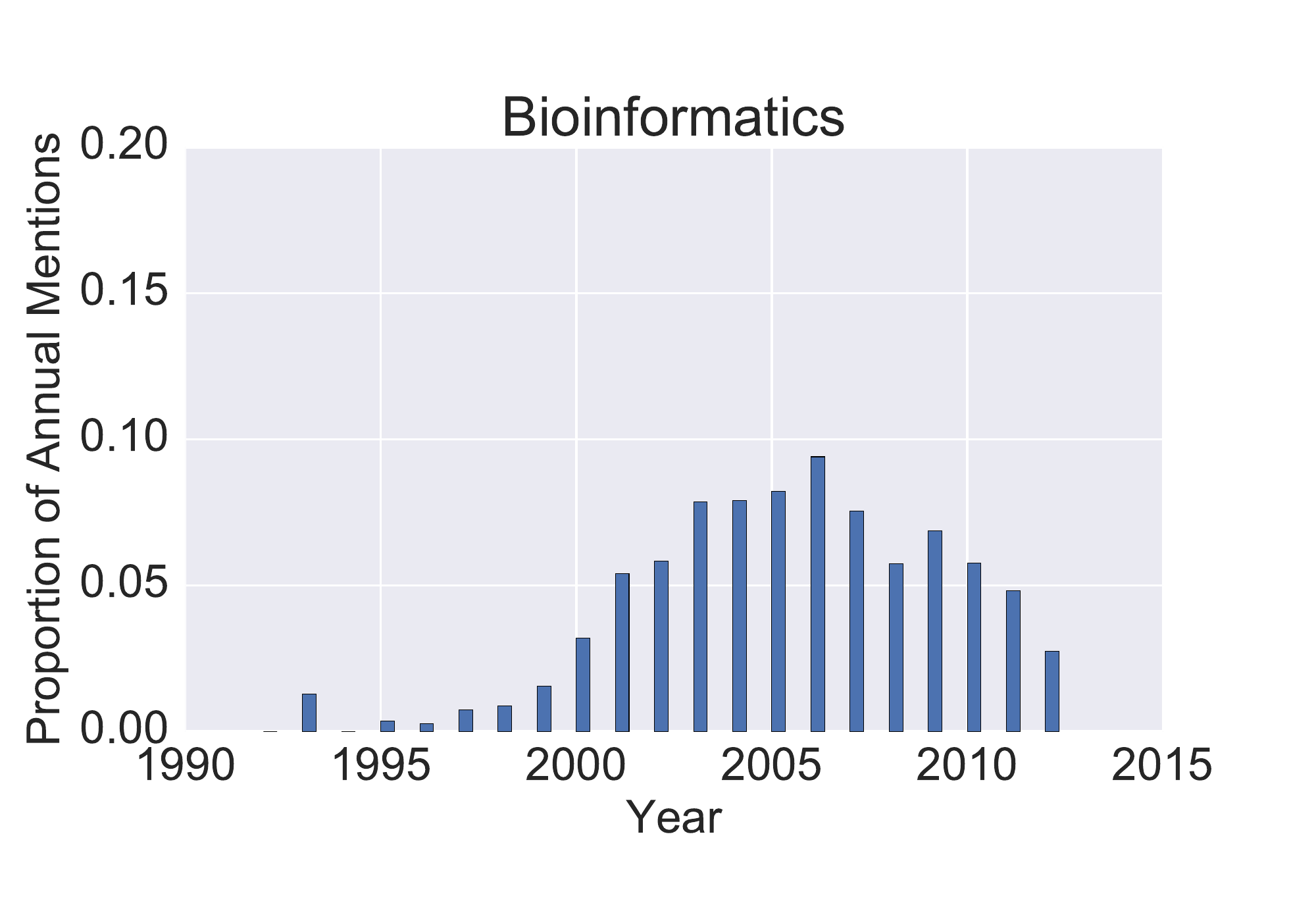}&
\includegraphics[width=.45\textwidth]{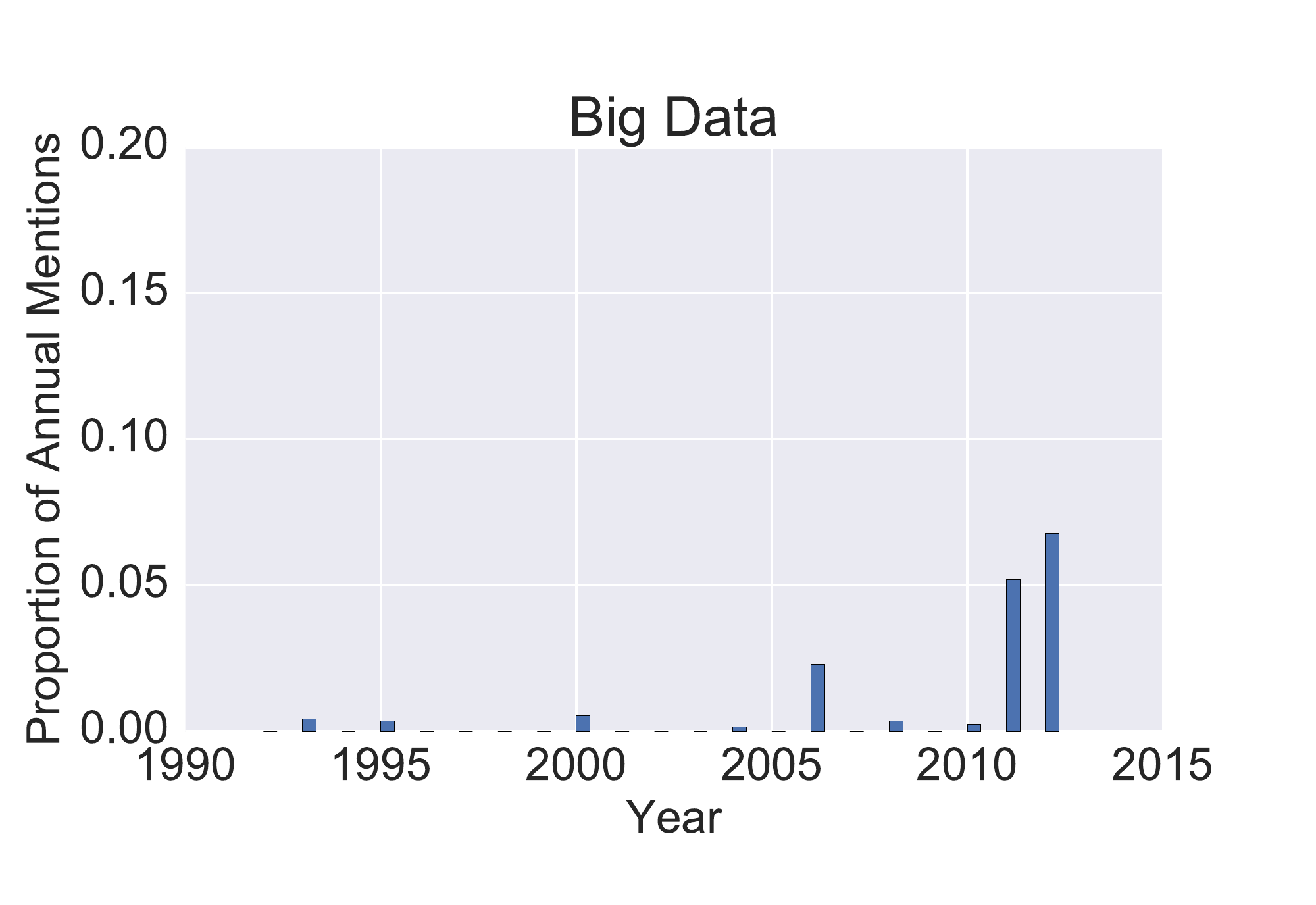}\\
\includegraphics[width=.45\textwidth]{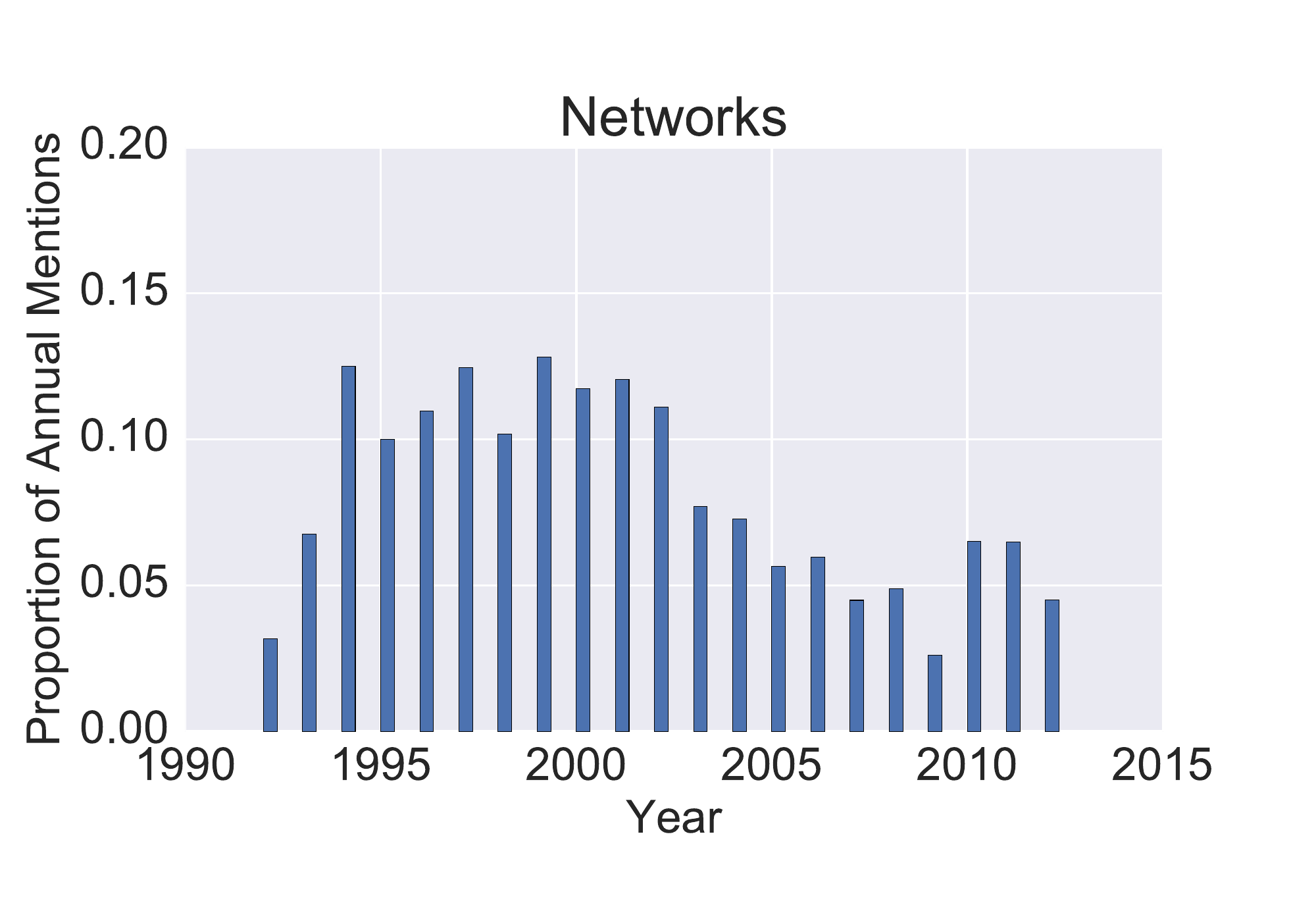}&
\includegraphics[width=.45\textwidth]{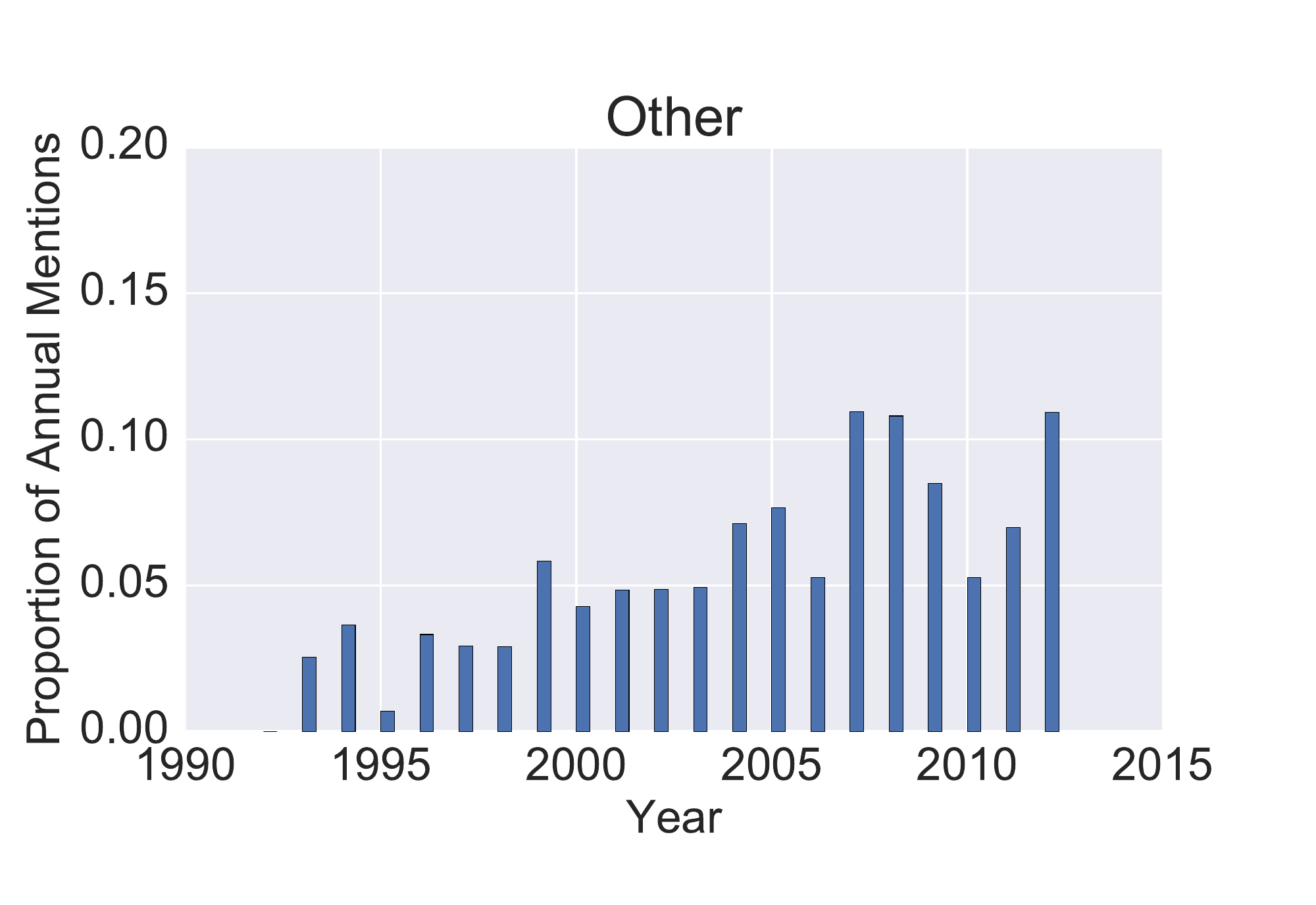}\\
\end{tabular}
\end{center}
\caption{Yearly proportion advertisements mentioning specific rooms in the game of informatics that are currently increasing or have peaked and declined in a relatively short time.}\label{fig:individuals2}
\end{figure}

Many individual topics that have substantial proportions of job advertisements are shown separated out from the set in Figures \ref{fig:declining} though \ref{fig:individuals2}.  In Figure \ref{fig:declining}, we see a set of areas that show a dramatic boost at the beginning of the 1990s and then a steady drop off over the period of 20 years.  Graphics peaked a little later, but still early in our dataset, and then steadily declines. 


\begin{figure}[t]
\begin{center}
\begin{tabular}{cc}
\includegraphics[width=.45\textwidth]{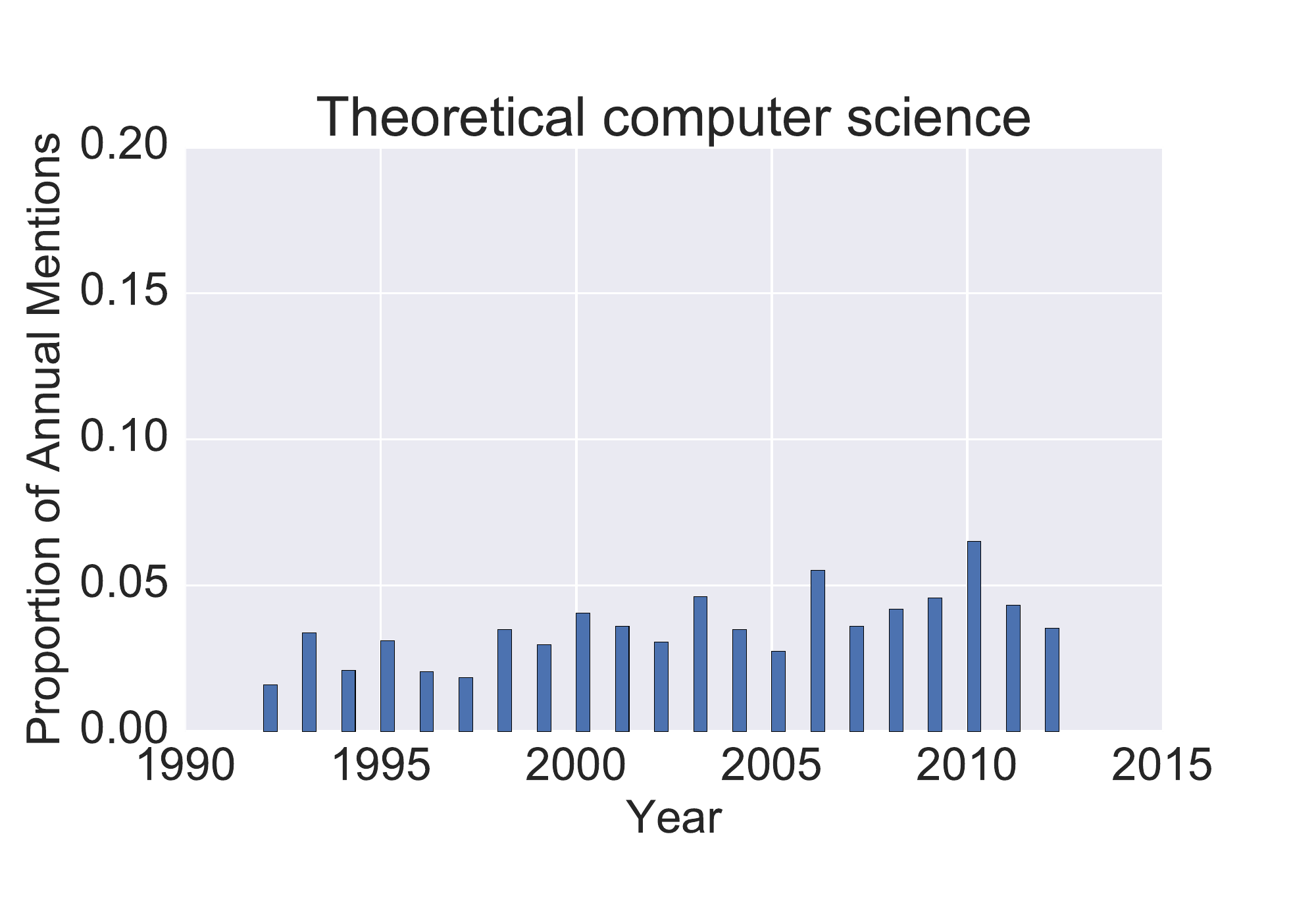}&\includegraphics[width=.45\textwidth]{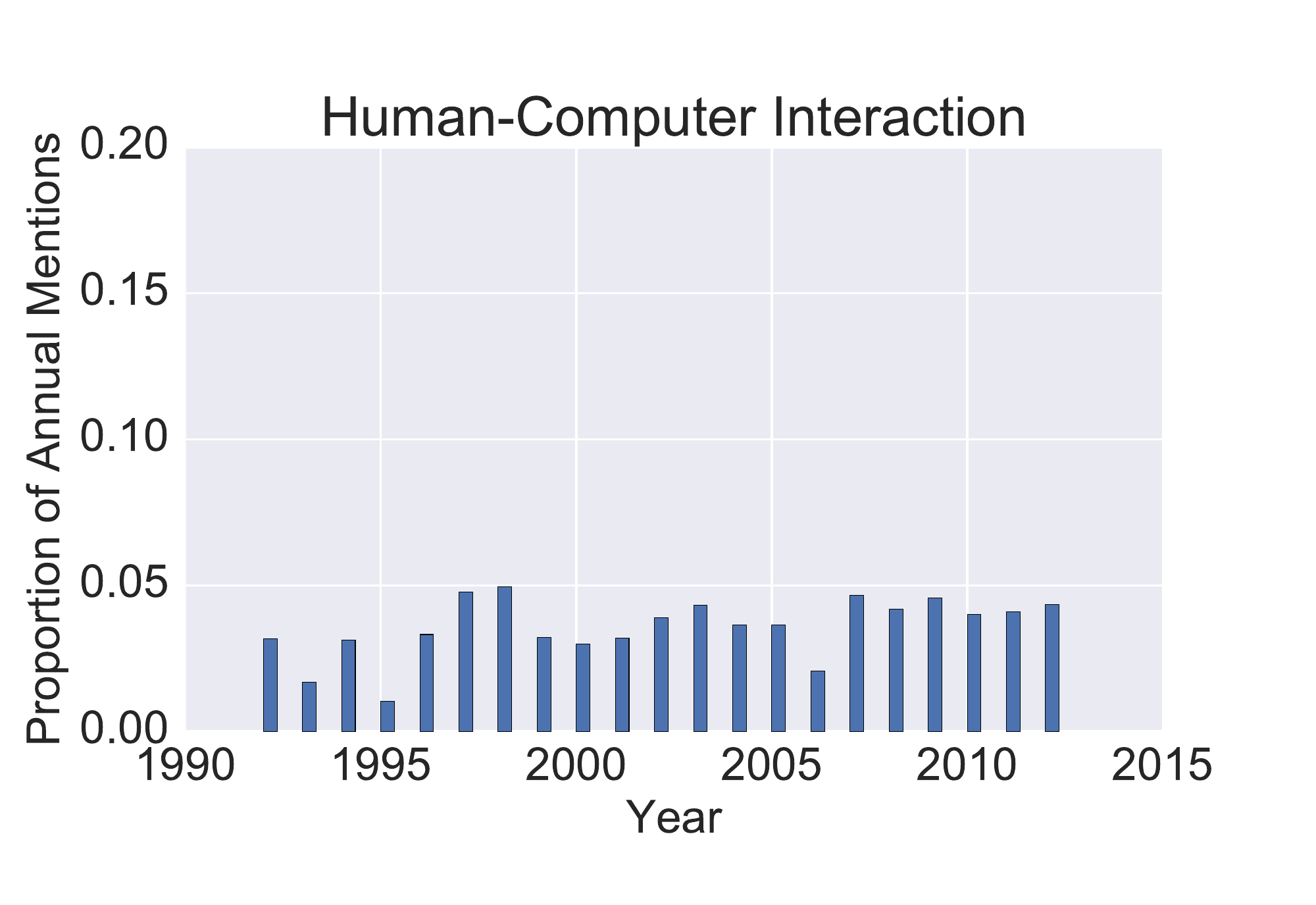}\\
\includegraphics[width=.45\textwidth]{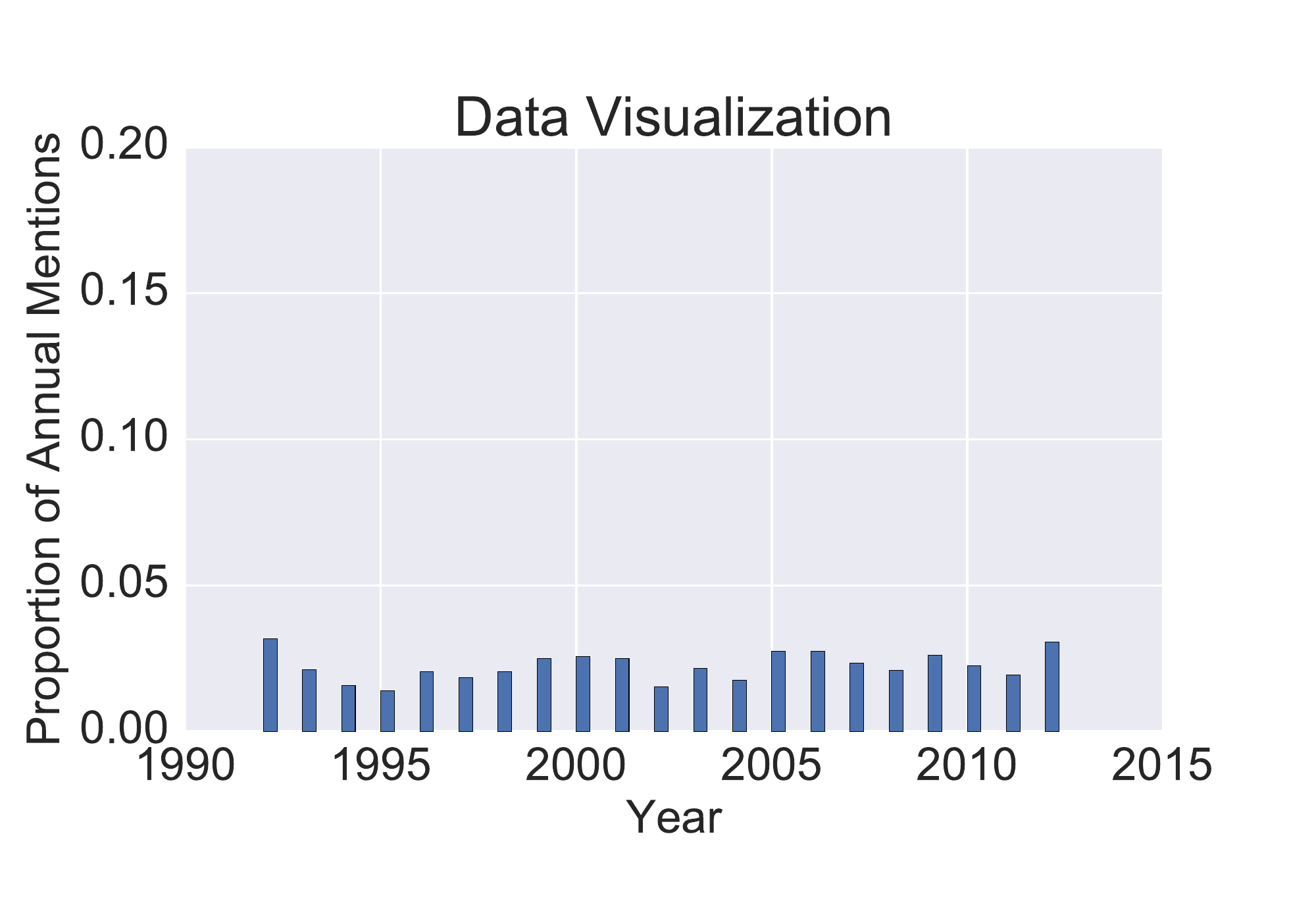}&
\includegraphics[width=.45\textwidth]{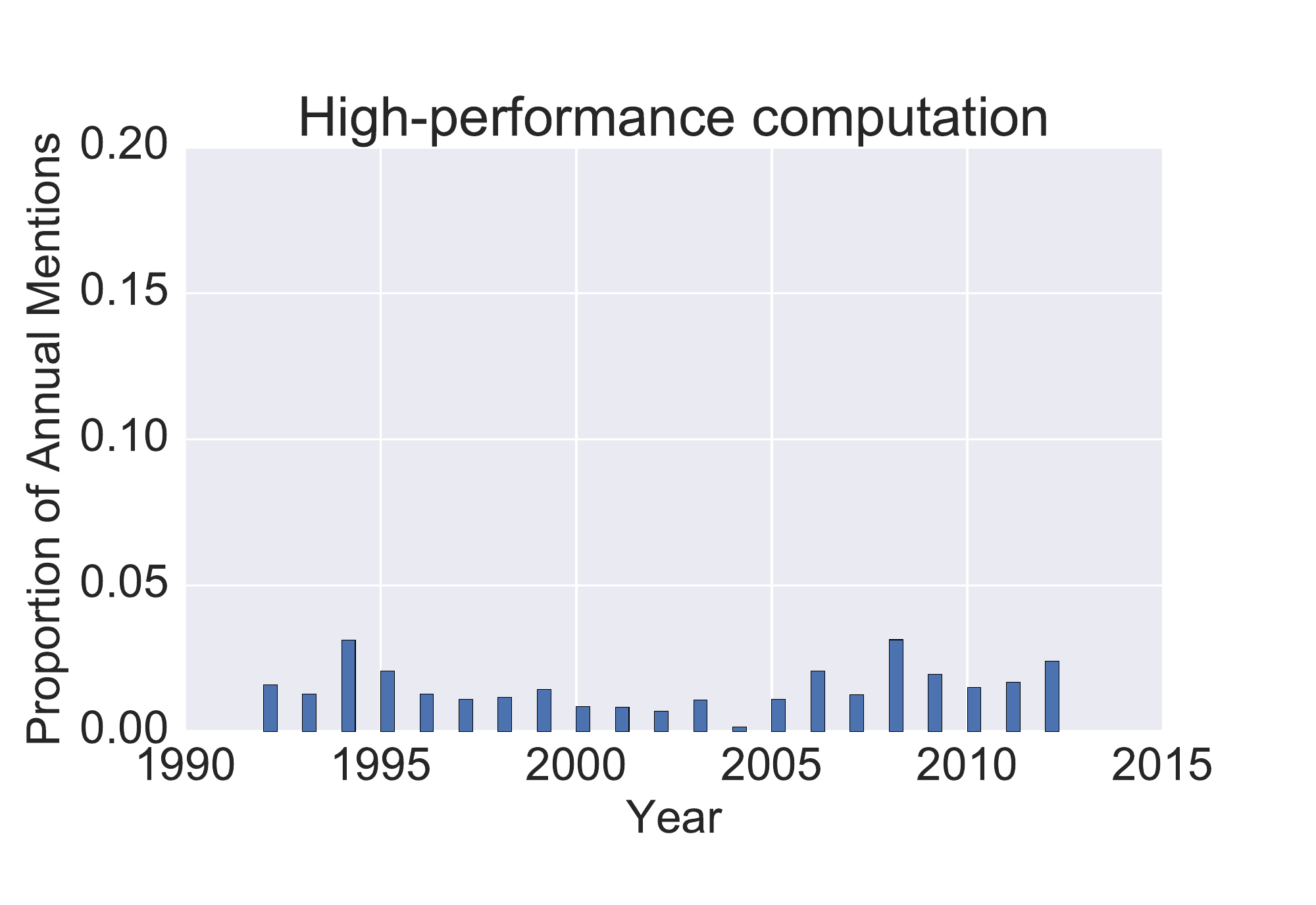}

\end{tabular}
\end{center}
\caption{Yearly proportion advertisements mentioning areas of Computer Science and Informatics that appear to have remained stable over a long period of time.}\label{fig:static}
\end{figure}

In Figure~\ref{fig:static} we see four areas that, aside from occasional peaks, are relatively stable and steady in their hiring.  It is interesting that in the case of Theoretical Computer Science (TCS), there is a slight increase over time though this may be an artefact of the code groups, with data structures and algorithms being particularly broad.

Figure~\ref{fig:individuals2} shows six areas that seem to still be climbing in hiring trends.  Security and Artificial Intelligence/Machine Learning are both enjoying healthy surges in hiring, and we have separated mentions of Big Data to show a particular field that has become popular in the last 5 years.  Finally, the Other category in Figure~\ref{fig:individuals2} is interesting, as it has substantially increased over the time period. This category includes several interdisciplinary topics such as e-commerce, arts and computing, heritage and computing etc., plus some other specialized subareas that do not fit into any of the 26 other topic areas.

\subsection{Discussion}

From the results above, there is an interesting story that emerges in relation to the \emph{Clue} hypothesis.  We see a large number of areas that have a spike in hiring and then a steady decrease to a consistent level of hiring.  These tend to be quite fundamental research areas to computer science --- when we were PhD students, Helmut would call these ``The very pipets and reagents we work with.''   While it is very likely that Theoretical Computer Science had an earlier peak, when giants strode boldly into new domains, the early 1990s had spikes of hiring in systems focussed foundational research.  Core systems topics were invested in first by departments, specifically Operating Systems, Parallel and Distributed Systems, Software Engineering, and Programming Languages.  Following these areas, we see the \emph{Clue}-like behaviour in Networks, Databases, and Computer Graphics, which lag behind with their peaks, and then follow the same drop off trend in hiring.  Younger fundamental areas, like HCI and Data Visualization, seem to miss this investment period, but have maintained a steady presence in hiring.  

The data seems to match societal shifts that were happening at the time.  This investment in core systems topics Computer Science by institutions coincides with the advent of reasonably inexpensive personal computers and just prior to the mainstreaming of graphical user interfaces and internet access.  Computing technology exploded into the mainstream consciousness of society, and that paved the way for investment in these foundations just prior to our dataset.  Given our sample starts at a notable peak in Computer Science research and practice, it is likely that we are seeing the end of a very long investigation of \emph{Clue} rooms that began in the 1970s with steady growth on the back of advances in so many technologies, from the Xerox Star to ARPANET, and then a steady drop-off to a base level for the above fundamental areas over 20 years.  

While we posit that some of this hiring is as a result of that increased availability of resources at institutions for Computer Science systems research as it became mainstream and impactful, with another contributing factor for being the academy reacting to a steady increase in student numbers.  After all, it would only be a few short years after these big periods of investment by institutions that the first dot-com boom and bust happened where teaching numbers bloomed, just as the authors were entering their PhD programmes.  

After the ascendency of these foundational areas, we can see Security surges in 2005, which corresponds to the first time over 50\% of the developed world had internet access and the rise of Google~\cite{ITU:2015}.  For AI/Machine Learning, which many would consider a foundational area, we seem to have captured the end of the AI Winter~\cite{Kurzweil:2005} with the steady climb in a variety of different contexts over the 20 years.  Each of these areas seem to be enjoying particularly long periods of targeted hiring.

The question remains: why has hiring gone down in foundational areas after the year 2000?  The answer likely lies in the increasingly large number of areas of investigation that emerged during the sample. Bioinformatics, Games, Web Technologies, and the myriad of research questions in the Other category, have led to an increasingly fragmented field as evidenced in Figure~\ref{fig:overallresults}.  That fragmentation tends to lend evidence to people dividing up into different rooms to ask questions for a period of time.  These rooms likely do contribute knowledge, to some extent, to mature foundational areas as well and to the game as a whole.  For example, when someone is hired into convergent areas such as bioinformatics, computational neuroscience, or digital heritage, there will be advances to the foundational areas in support of these new areas.  It is also likely the case that, as interdisciplinarity has grown, having people who are in areas that are malleable to application areas has been of benefit to departments as it potentially allows them to adapt to changing priorities in the research landscape, while still addressing the foundational teaching that can be delivered, decreasing the need for targeted hirings in those areas as the pioneers retire.  However, it is important to note that many of specific areas appear and nearly vanish from the landscape very quickly, giving weight to the \emph{Clue} analogy.  

One very interesting feature in the data is that, going back to 1992, there was a relatively low percentage of Open positions, ranging from approximately 10\% down to almost nothing in any given year.  This tends to indicate that academic institutions, in general, have always been quite specific in their Computer Science hiring, something that is often anecdotally attributed to modern hiring practices.  While the drivers behind hiring trends remain complex and difficult to interrogate, this contributes to the theory that informatics research is driven by specific needs at the time within the institutions and the funding landscape.  The analysis of the academic job advertisements does seem give strong evidence that the \emph{Clue}-like behaviour posited by Helmut is alive and well in the academic community.  

Now that we have established that this behaviour is indeed happening in the academy, we will investigate whether this is an optimal strategy for research.

\section{The Game of Informatics}

\subsection{Preliminaries}

Reflecting Helmut's broadly interdisciplinary approach to research, we now apply the model of Neal~\cite{neal1999} --- exposited originally in the quantitative economics literature to model the dynamics of career choice --- to investigating topic- and field-switching amongst researchers. We make several assumptions, chief amongst them the assumption that researchers will seek to optimize their level of grant funding.

Define total research grant income $g_t = \theta_t + \epsilon_t$ where $\theta_t$ is funding due to the choice of \textit{research field} while $\epsilon_t$ is funding due to choice of \textit{research topic}. We proceed with discrete time modelling. At each time point $t$, a researcher has three options: 
\begin{enumerate}
\item Change nothing
\item Select a new primary research topic, within current field ($\epsilon_t$ resampled)
\item Switch research fields (both $\theta_t$ and $\epsilon_t$ are resampled)
\end{enumerate}

Observed grant funding levels allocated by granting agencies typically assume a broad range of values, with the grants at the extrema being relatively less common. In our model, it is necessary to select, a priori, distributions $F$ and $G$ from which we may sample the funding level components $\theta_t$ and $\epsilon_t$ for each individual researcher. In our investigation here, we will begin by grossly oversimplifying reality and assuming that these distributions are uniform, but then move on to more realistic choices of $F$ and $G$ that match observed grant histograms for Canada's NSERC.

The game for each researcher now becomes one of maximizing lifetime grant income --- keeping in mind that our researchers are only permitted to annually select from the three choices described above. A very simple approach would be to simply optimize the sum of annual grant income over the number of research-active years of the researcher. A major challenge with this approach is that it assumes perfect optimism for a static future. A common approach to make a more realistic model is to include a ``future discounting'' factor ($\beta$) for which values less than one express correspondingly less certainty about the future.

We draw $\theta_t \sim F, \epsilon_t \sim G$, with $F$ and $G$ independent and denote the future discount as $\beta$. Our goal is to maximize
\begin{equation} \label{eq:1}
 \mathbb{E}\sum_{t=0}^{\infty} \beta^t g_t.
\end{equation}

Our value function $V(\theta,\epsilon)$ is the maximum of equation \ref{eq:1} over all feasible policies.
The maximum is found computationally by iterating to convergence on the Bellman equation\footnote{viz., $\mbox{max}( \theta + \epsilon + \beta V(\theta,\epsilon), \theta + \int \epsilon'G(d\epsilon') + \beta \int V(\theta,\epsilon') G(d\epsilon'), \int \theta' F(d\theta) + \int \epsilon'G(d \epsilon') + \beta \int \int V(\theta',\epsilon') G(d\epsilon) F(d\theta') )$}. We implemented this calculation directly with a simple Python script. A full description of this process is beyond the scope of this paper; the details are presented concisely in Neal~\cite{neal1999} and the interested reader may refer to \cite{RePEc:mtp:titles:0262018748} for a gentler exposition.

\subsection{Optimal policies}

We investigate optimal policies; $\theta$ and $\epsilon$ take values between 0 and 5 and, as per the exposition above, their sum represents the funding level of the researcher.

Consider first a grossly oversimplified situation where $F$ and $G$ are strictly uniform and researchers are optimistic about the future ($\beta = 0.95$). 

\begin{figure}
\begin{center}
\renewcommand{\arraystretch}{.5}
\begin{tabular}{cc}
\includegraphics[height=5.7cm]{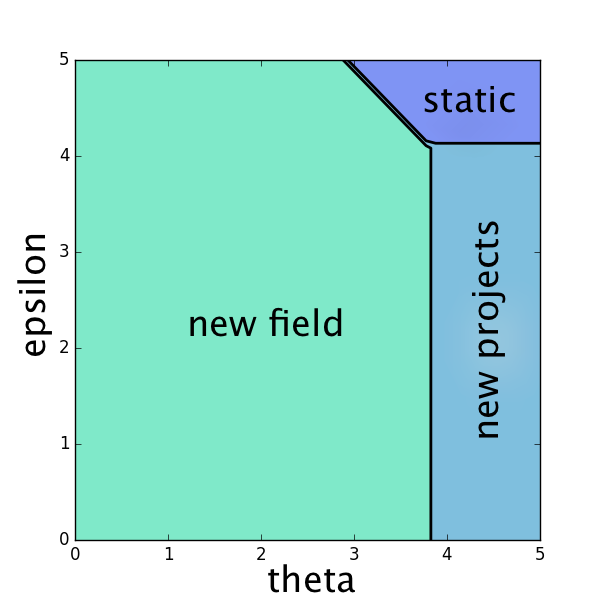}&
\includegraphics[height=5.7cm]{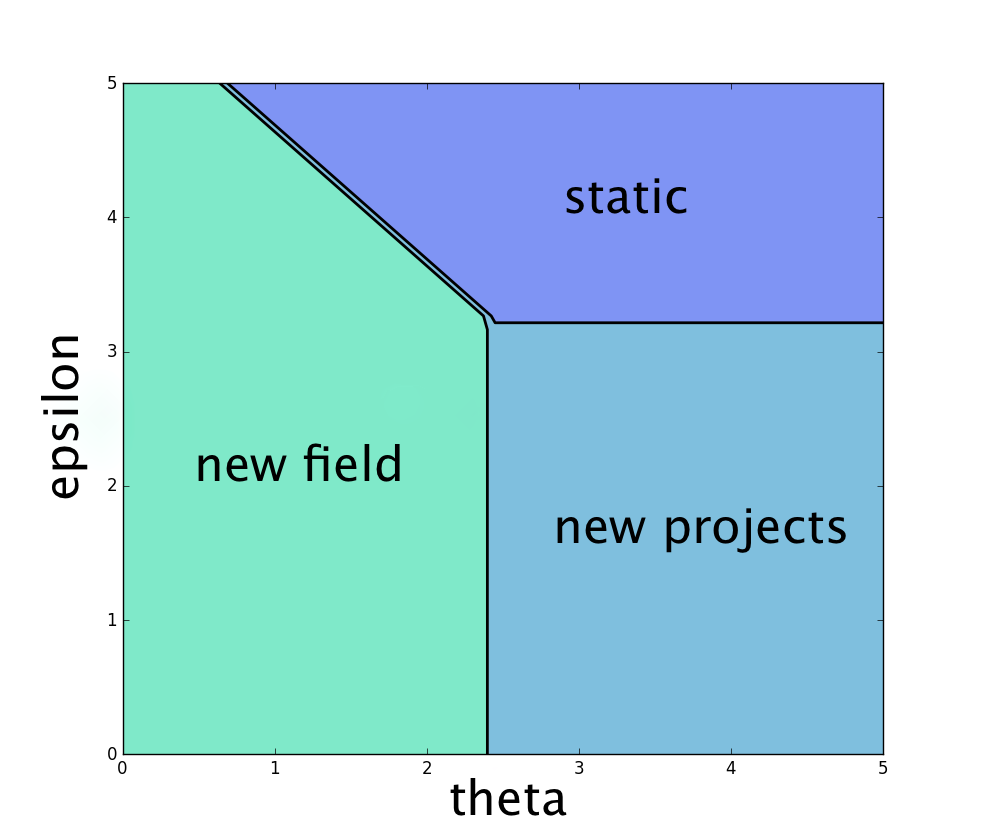}\\
(a) & (b)\\
\end{tabular}
\end{center}
\caption{Discrete modelling results (a) when $F$ and $G$ are strictly uniform, $\beta=0.95$; (b) when $F$ and $G$ are modified beta binomial, $\beta=0.95$. In (a), the policy space is dominated by the strategy of switching fields, whereas it is less so in (b).}\label{fig:simplots1}
\end{figure}

Under these simplified conditions, the optimal policy is dominated by encouraging researchers to switch fields until settling in a field with sufficiently high $\theta$, after which a researcher will experiment with new topics until reaching a static condition with high $\theta$ and $\epsilon$ (Figure \ref{fig:simplots1}(a)). Though this policy is certainly consistent with the observed \textit{Clue}-like behaviour discussed in this manuscript, the modelling assumptions are not particularly realistic.

We now consider a more realistic model in which the distributions $F$ and $G$ are synthetic bimodal distributions consisting of a modified beta binomial with varying parameters $a, b$ investigated, and an additional mass at zero. The mass at zero represents the risk of losing funding altogether when changing fields or topics. For $F$ we begin with a beta binomial having parameters $a=10.0,b=10.0$ and then add a mass with probability 0.6 at zero, to model the very real risk of losing funding when switching fields. For $G$ we choose a straight beta binomial having parameters $a=10.0,b=10.0$. The researchers remain optimistic with a future discount of $\beta = 0.95$.

Under these more complex conditions, we see that the impetus to switch fields occupies a smaller portion of the policy space (Figure \ref{fig:simplots1}(b)). The high risk of switching fields replaces a dominant field-switching strategy with a combination of a within-field topic switches and simply settling for what one already has.

If we fix $F$ and $G$ as above but our researchers develop pessimism about the future in a post-Trumpian dystopia --- modelled with a steep future discount $\beta=0.5$ --- we find, unsurprisingly, that field switching is quite often suboptimal.

\begin{figure}
\begin{center}
\includegraphics[height=5.7cm]{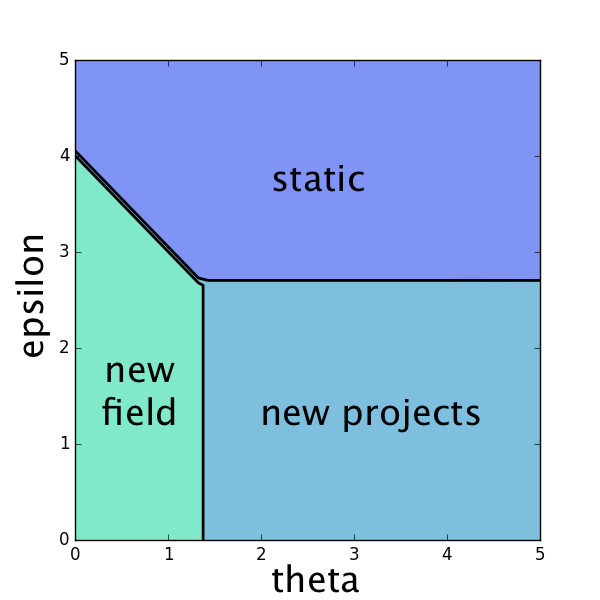}
\end{center}
\caption{Discrete modelling results when $F$ and $G$ are modified beta binomial, $\beta=0.5$.}
\end{figure}


If one is prepared to accept the assumptions of this model, one may conclude that in the presence of favourable conditions for research funding (and optimism that these conditions will continue), \textit{Clue}-like field switching is an optimal policy for all but the most senior, best funded, researchers.


%
%
%

\section{Conclusions}

A preliminary exploration of Helmut J\"urgensen's hypothesis on the
fast-changing nature of informatics research is presented.
For this purpose, first, a content analysis of university job advertisements
is performed for a 20 year period ending in 2012. Many areas do indeed
demonstrate a spike in hiring, followed by a steady decrease. Theoretical computer science remained relatively consistent throughout this entire period, which may be the end-result for the other foundational areas of computer science that have declined during this period. Other more specific areas seem to spike and descend to very low numbers quite rapidly.  Future work will apply statistical analysis to the determine the statistical significance of the observed trends.

Then, simulation techniques from quantitive economics were used. It was found that as optimism regarding funding levels increases, this seems to increase the pace of switching research areas, as an optimal strategy.

\biblio{helmut} 
\end{document}